\documentclass[prl,aps,twocolumn,floatfix,showpacs]{revtex4-1}
\usepackage{graphicx,graphics,psfrag,amsmath,calc}
\usepackage[dvips]{color}
\begin{document}
\title{Ultra-cold fermions in real or fictitious magnetic fields: \\
The BCS-BEC evolution and the type-I--type-II transition}
\author{M. Iskin$^1$ and C. A. R. S{\'a} de Melo$^2$}
\affiliation{$^1$Department of Physics, Ko{\c c} University, 
Rumelifeneri Yolu, 
34450 Sariyer, Istanbul, Turkey. \\
$^2$School of Physics, Georgia Institute of Technology, Atlanta, Georgia 30332, USA.}
\date{\today}

\begin{abstract}
We study ultra-cold neutral fermion superfluids 
in the presence of fictitious magnetic fields, as well as 
charged fermion superfluids in the presence of real magnetic fields. 
Charged fermion superfluids undergo a phase transition from type-I to type-II 
superfluidity, where the magnetic properties of the superfluid change 
from being a perfect diamagnet without vortices to a partial diamagnet
with the emergence of the Abrikosov vortex lattice. The transition from 
type-I to type-II superfluidity is tunned by changing the scattering parameter
(interaction) for fixed density.
%
%
We also find that neutral fermion superfluids such as $^6$Li and $^{40}$K 
are extreme type-II superfluids, and that they are more robust to 
the penetration of a fictitious magnetic field in the BCS-BEC 
crossover region near unitarity, where the critical fictitious magnetic 
field reaches a maximum as a function of the scattering parameter 
(interaction). 

\pacs{03.75.Ss, 03.75.Hh, 05.30.Fk}
\end{abstract}
\maketitle

%
%
A key experiment in the verification that neutral Fermi superfluids can evolve 
from the Bardeen-Cooper-Schrieffer (BCS) to the Bose-Einstein condensation 
(BEC) regime was the observation of quantized vortices throughout the BCS-BEC 
evolution~\cite{ketterle-2005} upon rotation of the atomic cloud.
This observation had a very dramatic impact beyond the atomic physics community, 
because it showed that superfluidity of Cooper pairs and of tightly 
bound bosonic molecules for s-wave pairing are the manifestation of 
the same type of physics.  The key tool that permitted such realization 
is the tunability of the interaction between fermions through the 
use of Feshbach resonances. The same kind of tunability does not exist 
in $^3$He, the standard condensed matter neutral superfluid, or in 
superconductors. The situation is even worse in neutron and proton superfluids, 
which are thought to exist in the core of neutron stars. 

Very recently, a new technique was developed that permitted
the production of fictitious magnetic fields which can couple 
to neutral bosonic atoms~\cite{spielman-2009a, spielman-2009b}. 
These fictitious magnetic fields are produced through an all 
optical Raman process, couple to a fictitious charge, but produce
real effects like the creation of vortices in the superfluid state of bosons. 
In principle, the same technique can be applied 
to ultra-cold fermions, which coupled with the control over interaction
using Feshbach resonances allows the exploration of superfluidity not only
as a function of interaction, but also as a function of fictitious 
magnetic field. It is in anticipation of similar experiments involving 
ultracold fermions that we address in this manuscript the effects 
of fictitious magnetic fields on fermion superfluids as a function 
of interaction.

\begin{figure} [htb]
\centerline{\scalebox{0.50}{\includegraphics{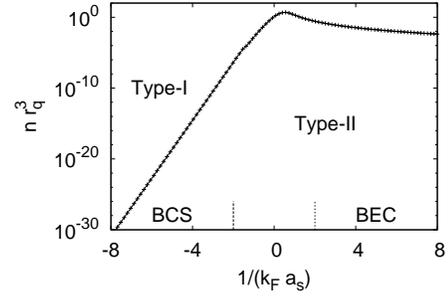} } }
\caption{ 
\label{fig:one}
Universal phase diagram of the dimensionless fermion density $n r_q^3$
versus scattering parameter $1/(k_F a_s)$, where $r_q = q^2/(m c_0^2)$
is the classical radius of a fermion with mass $m$ and charge $q$;
$c_0$ is the speed of light, $k_F$ is the Fermi momentum and $a_s$ is
the scattering length.  The dotted line separates regions of 
type-I and type-II superfluidity. 
}
\end{figure}

Unlike neutral superfluids, standard 
condensed matter charged superfluids (superconductors) can be of two 
types~\cite{abrikosov-1957}. 
Many superconductors are now known to be type-II 
(including heavy fermions, organics, and high-$T_c$ cuprates), 
where the application of an external magnetic field 
beyond the lower critical field $H_{c_1}$ leads to a non-uniform 
superfluid phase, which appears in the form of the Abrikosov 
vortex lattice, until a second critical field $H_{c_2}$ is reached,
when the system becomes normal. Other charged superfluids
are known to be type-I and do not allow the 
magnetic field to penetrate the sample. These systems are 
perfect diamagnets until the critical field $H_{c}$ is reached,
where the charged superfluid becomes normal. 
The parameter that characterizes the type of charged superfluid is the 
Ginzburg-Landau parameter $\kappa = \lambda/\xi$ corresponding to
the ratio between the penetration depth $\lambda$ of the magnetic field
into the sample and the coherence length $\xi$ of the charged superfluid,
such that type-I superfluids have $\kappa < 1/\sqrt{2}$ and
type-II have $\kappa > 1/\sqrt{2}$.

In this manuscript, we study neutral fermion superfluids in the presence
of fictitious magnetic fields and charged fermion superfluids in the 
presence of real magnetic fields as a function of interaction (scattering
parameter). We show that throughout the crossover region between BCS and BEC 
superfluidity both $^6$Li and $^{40}$K are extreme type-II superfluids, 
and for charged superfluids we find a phase transition from 
type-I to type-II superfluidity for fermions of density $n = k_F^3/(3\pi^2)$
interacting via a contact potential characterized by the 
interaction parameter $1/(k_F a_s)$. 
As shown in Fig.~\ref{fig:one}, the phase boundary in the density
$n$ versus interaction parameter $1/(k_F a_s)$ occurs 
when the critical value $\kappa_c = 1/\sqrt{2}$ is crossed. 
In the literature of charged superfluids the transition
from type-I to type-II was thought possible when induced by 
disorder and was described microscopically {\it only} in the 
BCS limit~\cite{abrikosov-1958, gorkov-1959, helfand-1966, abrikosov-1988}.
In contrast, here we show that, microscopically, a clean (no disorder) 
charged superfluid can exhibit a type-I to type-II transition induced 
by interactions. The phase diagram shown in 
Fig.~\ref{fig:one} has a wider applicability to include standard charged 
superfluids (like superconductors of condensed matter physics) and 
even proton superfluidity in nuclei or neutral stars, as long as the 
interactions can be described by a contact potential with corresponding 
scattering length $a_s$. In addition, we indicate that 
neutral (charged) superfluids are more robust 
to the penetration of ficitious (real) magnetic fields near unitarity, 
where the critical fictitious (real) magnetic fields reach a maximum as 
a function of the scattering parameter.

%
%
To describe the transition from type-I to type-II superfluidity as a function
of the interaction parameter and the
properties of neutral (charged) superfluids in the presence of
fictitious (real) magnetic fields during the BCS-BEC evolution 
for s-wave superfluids in three dimensions, 
we start with the Hamiltonian density
\begin{equation}
\label{eqn:hamiltonian-real-space}
{\bar {\cal H} } ({\bf r}) = \sum_\sigma
\psi^{\dagger}_{\sigma} ({\bf r}) 
\left( -\frac{ \hbar^2 \nabla^2}{2m} - \mu \right) 
\psi_\sigma ({\bf{r}})
+ {\hat U} ({\bf r}),
\end{equation}
where 
$
{\hat U} ({\bf r}) =  
\int d{\bf r'}
V ({\bf r}, {\bf r'}) 
\psi^{\dagger}_\uparrow ({\bf r})
\psi^{\dagger}_\downarrow ({\bf r'})
\psi_{\downarrow}({\bf r'})
\psi_{\uparrow} ({\bf r})
$
contains the attractive contact interaction potential 
$V ({\bf r}, {\bf r'}) = - g \delta ({\bf r} - {\bf r'})$,
and $\psi^{\dagger}_{\sigma} ({\bf r})$ is the creation operator 
of fermions with mass $m$ and spin $\sigma$. Notice that $g$ has dimensions
of energy times volume. To make progress, we rewrite 
the Hamiltonian ${\cal H} = \int d{\bf r} {\bar {\cal H}} ({\bf r})$ 
from real space to momentum space 
\begin{equation}
\label{eqn:hamiltonian-momentum-space}
{\cal H} = 
\sum_{ {\bf k}, \sigma } 
\xi_{\bf k} 
\psi^{\dagger}_{ {\bf k}, {\sigma}} \psi_{ {\bf k}, {\sigma}} 
- g \sum_{{\bf k}, {\bf k'}, {\bf q}}
 b^\dagger_{ {\bf k}, {\bf q} } b_{ {\bf k'}, {\bf q} }, 
\end{equation}
where 
$
b_{ {\bf k}, {\bf q} }^\dagger = 
\psi^{\dagger}_{ {\bf k} + {\bf q}/2, \uparrow }
\psi^{\dagger}_{ -{\bf k} + {\bf q}/2, \downarrow }
$
creates a fermion pair with center of mass momentum ${\bf q}$ and relative
momentum $2 {\bf k}$, $\xi_{\bf k} = \epsilon_{\bf k} - \mu$ is the kinetic
energy term with $\epsilon_{\bf k}  = \hbar^2k^2/(2m)$ 
and $\mu$ is the chemical potential.

%
Integration over the fermion fields~\cite{sademelo-1993} leads to 
the order parameter equation
\begin{equation}
\label{eqn:order-parameter}
\frac{1}{g} 
= 
\frac{1}{L^3}
\sum_{\bf k} 
\frac 
{\tanh 
\left[
\xi_{\bf k}/(2 T_c)
\right]
}
{ 2 \xi_{\bf k} } 
\end{equation}
at the critical temperature $T_c$, where the order parameter vanishes. 
Here $L^3$ is the sample volume.
The interaction $g$ can be written in terms of the scattering length $a_s$ 
leading to 
$
1/g 
= 
- 
m/
\left( 
4\pi \hbar^2 a_s 
\right)
+ 
(1/L^3)
\sum_{\mathbf{k}} 
\left[ 
1/( 2 \epsilon_{\bf k} )
\right]$.
The second self-consistency relation is the number equation
\begin{equation}
\label{eqn:number}
N 
= 
\sum_{ {\bf k}, \sigma } 
{\rm f} (\xi_{\bf k}) 
+ 
T_c \sum_{q} 
\frac{ \partial \left[ \ln ( L^3 {\cal K} / T_c ) \right] } { \partial \mu }
\end{equation}
where ${\rm f} (\xi_{\bf k})$ is the Fermi function, and
\begin{equation}
\label{eqn:pair-propagator}
{\cal K}^{-1} = 
\frac{1}{g} - 
\frac{1}{L^3}
\sum_{\bf k} 
\frac { 
1 
- {\rm f} ( \xi_{ {\bf k} + {\bf q}/2 } ) 
-  {\rm f} (\xi_{ -{\bf k} + {\bf q}/2}  ) 
}
{ \xi_{ {\bf k} + {\bf q}/2 } + \xi_{ -{\bf k} + {\bf q}/2}  - i \hbar \omega }  
\end{equation}
is the pair propagator, and $\omega$ is the Matsubara frequency for bosons.

%
%
The effective action is 
$
T S_{\rm eff}/\hbar = 
\sum_q {\cal K}^{-1} (q) 
\vert \Delta (q) \vert^2 
+ \frac{b}{2 L^3}
\sum_{q_1, q_2, q_3}  
\Delta (q_1) \Delta^* (q_2) \Delta (q_3) \Delta^* (q_1 - q_2 + q_3) 
$
in terms of the  order parameter $\Delta (q)$,
where $q = ({\bf q}, i\omega)$. 
To study thermodynamic properties, 
we take $i \hbar \omega = 0$, or equivalently, 
$\Delta ({\bf r}, \tau) \equiv \Delta ({\bf r})$, 
leading to the effective Lagrangian 
density
$$
{\cal L}_{\rm eff} = 
 a \vert \Delta \vert^2 
+ \sum_{i,j} \frac{\hbar^2 c_{ij}}{2m} \nabla_i {\bar \Delta} \nabla_j \Delta
+ \frac{b}{2} \vert \Delta \vert^4. 
$$
Using the notation $X_{\bf k} = \tanh[\xi_{\bf k}/(2T)]$ and 
$Y_{\bf k} = {\rm sech}^2 [\xi_{\bf k}/(2T)]$,
the coefficients of the Lagrangian 
density are
$
a (\mu, T) 
= 
\frac{1}{g} - 
\frac{1}{L^3}
\sum_{\bf k} 
\frac{X_{\bf k}} { 2\xi_{\bf k} } 
$
for the constant term,
$$
L^3 c_{ij} (\mu_c, T_c) =  
\sum_{\bf k}
\left[
\left(
\frac{X_{\bf k}}{8\xi_{\bf k}^2}
- \frac{Y_{\bf k}}{16\xi_{\bf k} {T_c}}
\right)
\delta_{ij} 
+ \frac{X_{\bf k} Y_{\bf k}}{T_c^2} \frac{\hbar^2 k_i k_j} {16 m \xi_{\bf k} }
\right]
$$
for the coefficient of the gradient terms 
$\nabla_i {\bar \Delta} \nabla_j \Delta,$ 
and
$
L^3 b (\mu_c, T_c) =  
\sum_{\bf k} 
\left( \frac{X_{\bf k}}{4 \xi_{\bf k}^3 } - 
\frac{Y_{\bf k}}{ 8\xi^2_{\bf k} T_c} \right)
$
for the coefficient of the non-linear quartic term.
Notice that $c_{ij} =  c \delta_{ij}$ for s-wave superfluids.
%
%

In general, near $T_c$,
$
a(\mu, T) 
= 
-
a_0 
\epsilon (T),
$ 
where 
$
a_0 
= 
T_c 
\left[ 
\partial a
/
\partial T
\right]_{T_c}
$
and 
$\epsilon (T) = (1 - T/T_c)$.
In the BCS limit of $1/(k_F a_s) \to -\infty$, 
$
L^3 a_0 
= 
{\cal D}_F,
$
where
$
{\cal D}_F 
= 
m k_F L^3/(2\pi^2 \hbar^2)
$
is the density of single particle states per spin channel 
at the Fermi energy $\epsilon_F$.
Also the coefficient of the quartic term is 
$
L^3 b 
= 
\left[ 
7 \zeta(3)/(8 \pi^2 T_c^2)
\right]
{\cal D}_F ,
$
while the coefficient of the gradient term is
$
L^3 c 
= 
\left[
7 \zeta(3)/(12 \pi^2 T_c^2) 
\right]
{\cal D}_F 
\epsilon_F.
$
Here, the zeta function $\zeta(3) = 1.202$, while
the critical temperature 
$
T_c 
= 
\left(
8 e^{\gamma-2}/\pi
\right)
\epsilon_F
\exp
\left[
-\pi/(2 k_F \vert a_s \vert)
\right],
$
with $e^\gamma \approx 1.781$,
and the chemical potential $\mu = \epsilon_F$.
However, in the BEC limit of $1/(k_F a_s) \to + \infty$,
$
L^3 a_0 
=
{\cal D}_F \epsilon_F
/
(4 \vert \mu \vert).
$
Correspondingly the coefficient of the quartic term is
$
L^3 b
=
\left(
\pi/32
\right)
{\cal D}_F 
/
(
\vert \mu \vert \sqrt{\epsilon_F \vert \mu \vert}
),
$
and the coefficient of the gradient term
is
$
L^3 c 
=
\left(
\pi/16
\right)
{\cal D}_F 
/
\sqrt{\epsilon_F \vert \mu \vert}.
$
In this case, $T_c \approx 0.218 \epsilon_F$
and 
$
\mu 
= 
E_b/2, 
$
where 
$
E_b 
= - \hbar^2/(m a_s^2)
$
is the two-particle binding energy in vacuum.

Next, we scale the order parameter 
to $\psi ({\bf r}) = \sqrt{c} \Delta ({\bf r})$ and introduce 
an external (real or fictitious) 
magnetic field via the vector potential ${\bf A}({\bf r})$,
using the substitution $\nabla_i \to \nabla_i - 2iq A_i/(\hbar c_0)$,
where $q$ is the real or fictitious particle charge 
and $c_0$ is the speed of light.
The difference in free energy density between 
the charged superfluid and its normal state in the presence of 
magnetic fields takes the Ginzbug-Landau form
$$
{\cal F}_{GL} = 
\alpha \vert \psi \vert^2
+
\frac{\beta}{2} \vert \psi \vert^4
+
\frac{\hbar^2}{2 m}
\left|
\left(
-i \nabla - \frac{2q}{\hbar c_0} {\bf A}
\right)
\psi
\right|^2	
+
\frac{\vert {\bf H} \vert^2}{8\pi}
$$
where ${\bf H} = \nabla \times {\bf A}$ is the real or fictitious microscopic 
magnetic field. The parameter $\alpha = a/c$ changes
sign at $T = T_c$, 
%
however $\beta = b/c^2$ is always positive guaranteeing the stability
of the theory. It is also useful to define the
flux quantum $\Phi_q = \pi \hbar c_0/q$, which will be used below.  

Minimization of ${\cal F}_{GL}$ with respect to $\psi$ and ${\bf A}$
lead the order parameter equation
\begin{equation}
\label{eqn:GL-order-parameter}
\alpha \psi 
+ 
\beta \vert \psi \vert^2 \psi 
+
\frac{\hbar^2}{2m} 
\left( 
-i\nabla - \frac{2q}{\hbar c_0} {\bf A} 
\right)^2 
\psi
=
0
\end{equation}
and to the current density
\begin{equation}
\label{eqn:GL-current}
{\bf j}
=
- \frac{\hbar q}{im}
\left(
\psi^* \nabla \psi - \psi \nabla \psi^*
\right)
-
\frac{4q^2}{mc_0} 
\vert \psi \vert^2 {\bf A}.
\end{equation}
Using the relation $\nabla \times {\bf H} = 4\pi {\bf j}/c_0$ and taking the 
curl of the current density leads to the London equation
$
\lambda^2 
\nabla 
\times 
\left(
\nabla \times {\bf H} 
\right)
+ {\bf H}
=0
$
where 
$
\lambda 
= 
\sqrt{ m c_0^2/(16\pi q^2 \vert \psi \vert^2) }
$
is the magnetic penetration depth.
Since 
$
\vert \psi \vert^2 
= 
\vert \alpha \vert/\beta 
= 
\vert a \vert c/b $ in weak magnetic fields,
the penetration depth becomes 
$
\lambda (T)
= 
\lambda_{GL}\vert \epsilon (T) \vert^{-1/2},
$
where 
$
\lambda_{GL} 
= 
\sqrt{b/(16\pi r_q a_0 c)}.
$
Here, $r_q = q^2/(m c_0^2)$ is
the classical radius of a fermion with mass $m$ and charge $q$ in CGS units.
%
%
Since $\vert \psi \vert^2$ plays the role of the superfluid density $n_s$,
we may write $\vert \psi \vert^2 = \vert \psi_0 \vert^2 \vert \epsilon (T) \vert
= n_s = n_{s,0} \vert \epsilon (T) \vert$, where
$n_{s,0} = \vert \psi_0 \vert^2 = a_0 c/b$ is the temperature independent 
prefactor. This observation allows us to write
$
k_F \lambda_{GL} =
\sqrt{
\left[ 3\pi/(16 k_F r_q) \right]
\left( n/n_{s,0} \right)
}.
$
The prefactor $n_{s,0}$ reflects a zero temperature extrapolation of the 
superfluid density $n_s$, however, in a Galilean invariant
system we must have $n_{s,0} \approx n/2$. 
Indeed, in the BCS limit $n_{s,0} = n/2$
such that $k_F \lambda_{GL} = \sqrt{3\pi/(8 k_F r_q)}$,
while in the BEC limit
$n_{s,0} = 3n/8$ leads to $k_F \lambda_{GL} = \sqrt{\pi/(2 k_F r_q)}$. 
The BCS value of
$k_F \lambda_{GL}$ is slightly smaller than its BEC value, 
however throughout the BCS-BEC evolution, $k_F\lambda_{GL}$ 
does not change substantially.

The coherence length can be extracted from Eq.~(\ref{eqn:GL-order-parameter}) as 
$
\xi(T) 
=
\hbar/\sqrt{
2 m \vert \alpha (T) \vert
}
$
leading to 
$
\xi (T) 
= 
\xi_{GL} \vert \epsilon (T) \vert^{-1/2}
$
where 
$
\xi_{GL} 
=
\hbar \sqrt{c/(2 m a_0)}.
$
Unlike the penetration depth, the coherence length $\xi_{GL}$ changes 
substantially during the BCS to BEC evolution.  
In the BCS regime,
$
k_F \xi_{GL} 
= 
\sqrt{7 \zeta(3)/(12\pi^2)} (\epsilon_F/T_c)
$
is very large, and in terms of $k_F a_s$ becomes
$
k_F \xi_{GL} 
=
A \exp [\pi/(2 k_F \vert a_s \vert)],
$
where 
$ 
A 
=
\sqrt{7 \zeta(3)/(12 \pi^2)}
(\pi e^{2-\gamma}/8).
$
In the BEC regime,
$
k_F \xi_{GL} 
= 
\sqrt{\pi/4}
\left( \vert \mu \vert /\epsilon_F
\right)^{1/4}
$
is also very large, and in terms of $k_F a_s$ becomes
$
k_F \xi_{GL} 
= 
\sqrt{\pi/4}/\sqrt{k_F a_s}.
$
However, $k_F \xi_{GL}$ passes through a minimun 
in the intermediate regime, where $k_F \xi_{GL} \approx {\cal O}(1)$.

\begin{figure} [htb]
\centerline{\scalebox{0.50}{\includegraphics{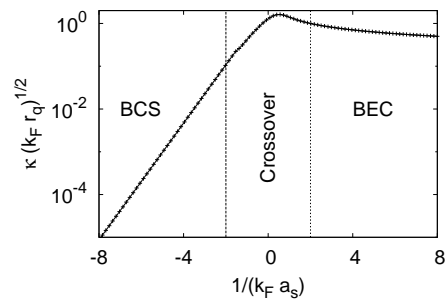} } } 
\caption{
\label{fig:two}
Universal plot of the Ginzburg-Landau parameter $\kappa$ 
versus scattering parameter $1/(k_{F} a_{s})$, where $r_q$ is the 
classical radius of a fermion with mass $m$ and charge $q$.
}
\end{figure}

As discovered by Abrikosov~\cite{abrikosov-1957}, the parameter 
$\kappa = \lambda (T)/\xi (T)$ is of fundamental importance in 
the characterization of the magnetic properties of charged superfluids. 
When $\kappa < \kappa_c = 1/\sqrt{2}$ the charged superfluid is
a perfect diamagnet (type-I), which does not allow the magnetic field to 
penetrate. When $\kappa > \kappa_c$, the 
charged superfluid allows the penetration of magnetic field
in the superfluid state in the form of vortices (type-II). 
Since the temperature dependence of $\lambda (T)$ and $\xi (T)$ is exactly the
same, the parameter 
\begin{equation}
\label{eqn:kappa}
\kappa 
=
\frac{\lambda (T)}{\xi (T)}
=
\frac{\lambda_{GL}}{\xi_{GL}}
=
\sqrt{
\frac{ m b}{8\pi r_q \hbar^2 c^2}
}
\end{equation}
is independent of temperature.
Notice that $\kappa \sqrt{k_F r_q} = \sqrt{s/(16\pi)}$,
where $s = k_F^3 b/(\epsilon_F c^2)$ is a dimensionless
parameter which is a function of $1/(k_F a_s)$ only.
In Fig.~\ref{fig:two}, we show the evolution of $\kappa$ as 
a function of the scattering parameter $1/(k_F a_s)$.
In the BCS limit, 
$
\kappa \sqrt{k_F r_q}
= 
\sqrt{
9 \pi^3/\left[14 \zeta(3)\right]
}
(T_c/\epsilon_F)
$
corresponds to 
$
\kappa \sqrt{k_F r_q}
= 
B \exp [- \pi/(2k_F \vert a_s \vert ) ],
$
where
$
B
=
\sqrt{
9 \pi^3/\left[14 \zeta(3)\right]
}
(
8 e^{\gamma-2}/\pi
).
$
While in the BEC limit, 
$
\kappa \sqrt{k_F r_q}
= 
\sqrt{2}
\left( \epsilon_F
/ 
\vert \mu \vert \right)^{1/4}
$ 
corresponds to $\kappa \sqrt{k_F r_q} = \sqrt{2} \sqrt{k_F a_s}$.
Notice the maximum of $\kappa$ in the vicinity of unitarity and 
$\mu = 0$ $(1/[k_F a_s] = 0.554)$. 

Unfortunately, in current experiments for neutral atoms in 
fictitious fields, only the product $q H$ is controlled, 
instead of $H$ alone~\cite{spielman-2009a}. 
So it is useful to think of ultra-cold superfluids like
$^6$Li or $^{40}$K as having charge $q \to 0$, but with $q H$ finite.
In this sense, these neutral superfluids are extreme
type-II with $\kappa \to \infty$ throughout the BCS to BEC 
evolution~\cite{footnote}.

To obtain the phase diagram shown in Fig.~\ref{fig:one}, we set
$\kappa = \kappa_c$ in Eq.~(\ref{eqn:kappa}) and 
extract the fermion density $n$ as a function of $1/(k_F a_s)$, 
which leads to $n  r_q^3 = s^3/(1536\pi^5)$.
In Fig.~\ref{fig:one}, $\kappa$ 
is higher (lower) than $\kappa_c$ below (above) the critical 
line indicating a type-II (type-I) charged superfluid phase.
For fixed density $n$, a phase transition from type-I to 
type-II charged superfluid occurs, as the interaction parameter 
$1/(k_F a_s)$ increases. 
Electron superfluids with 
$10^{21}~{\rm cm}^{-3} \le n \le 10^{23}~{\rm cm}^{-3}$ have  
$2.24 \times 10^{-27} < n r_q^3 < 2.24 \times 10^{-25}$,
and the transition between type-I and type-II occurs in the
interval $-8 < 1/(k_F a_s) < -4$, while 
proton superfluids in nuclear matter,
with $r_q^3 \approx 3.60 \times 10^{-48}~{\rm cm}^{-3}$ and 
$10^{37}~{\rm cm}^{-3} \le n \le 10^{38}~{\rm cm}^{-3}$,
have a type-I to type-II transition in the range $-4 < 1/(k_F a_s) < 0$. 

For type-I charged superfluids there is only the 
thermodynamic critical field $H_c (T)$ determined by the 
condition $H_c^2 (T)/(8\pi) = F_n  - F_s$, where $F_n$ $(F_s)$ is the Helmholtz
free energy for the normal (superfluid) state. For a uniform superfluid 
state the energy difference is $F_n - F_s = \alpha^2/(2\beta) = a^2/(2b)$
leading to $H_c (T) = H_{c,0} \vert \epsilon (T) \vert
= \vert a (T) \vert \sqrt{4\pi/b} $, where
$H_{c,0} = a_0 \sqrt{4\pi/b}$ is the temperature 
independent prefactor. 
%
%
We define the dimensionless thermodynamic critical field 
$\widetilde H_{c,0} = H_{c,0}/H_{k_F}$, where $H_{k_F} = \Phi_q k_F^2$. 
Notice that $\widetilde H_{c,0} = \hbar \omega_{c,0}/(2\pi \epsilon_F)$, where 
$\omega_{c,0} = \vert q \vert H_{c,0}/(mc_0)$ is the cyclotron frequency at
$H_{c,0}$. 
%
%
Using the asymptotic expressions for $a_0$ and $b$, we obtain
$
\widetilde H_{c,0} 
=
\sqrt{k_F r_q} 
\sqrt{4/[ 7\pi \zeta(3)]}
(T_c/\epsilon_F)
$
in the BCS regime, 
which can be rewritten as
$
\widetilde H_{c,0} 
=
\sqrt{k_F r_q} 
C
\exp 
\left[
- \pi/(2 k_F \vert a_s \vert)
\right],
$
with
$
C
=
\sqrt{4/[ 7 \pi \zeta(3)]}
(8 e^{\gamma-2}/\pi).
$
While we obtain
$
\widetilde H_{c,0} 
=
\sqrt{k_F r_q} 
(1/\pi^2)
(\epsilon_F/\vert \mu \vert)^{1/4}
$
in the BEC regime,
which can be rewritten as
$
\widetilde H_{c,0} 
=
\sqrt{k_F r_q} 
(1/\pi^2)
\sqrt{k_F a_s}.
$
The field 
$
\widetilde H_{c,0} 
$
reaches a maximum near unitary and $\mu = 0$, thus indicating that
type-I superfluids are most robust to the 
penetration of magnetic fields in that same region.

For type-II superfluids there are two 
critical fields. The first is called $H_{c_1} (T)$ and 
separates the perfect-diamagnet Meissner 
phase from the non-uniform phase exhibiting vortices.
The second is called $H_{c_2} (T)$ and separates
the non-uniform phase exhibiting vortices
from the normal state.
Since $^6$Li and $^{40}$K are extreme type-II superfluids
with $\kappa \to \infty$, then $H_{c_1} (T) \to 0$, and thus
we concentrate on the results for $H_{c_2} (T)$.
The calculation of $H_{c_2} (T)$ is performed
by linearizing Eq.~(\ref{eqn:GL-order-parameter}) 
$
- \hbar^2 
\left(
\nabla
- 
i
2\pi {\bf A}/\Phi_q
\right)^2
\psi
+ 
2 m \alpha (T) \psi
=
0.
$
Using the Landau gauge ${\bf A} = Hx \hat{\bf y}$, the momentum components
$k_y$ and $k_z$ are good quantum numbers and the solution for $\psi$ becomes
$\psi_{n, k_y, k_z} (x, y, z) = e^{(ik_y y + ik_z z)} u_n (x)$, which substituted
in the previous equation leads to the one-dimensional 
{\it Schr\"odinger} equation
$
\left[
- \hbar^2/(2m) d^2/dx^2 +
m \omega_s^2 (x - x_0)^2/2 
\right]
u_n (x) 
= 
\epsilon_n u_n (x),
$
where $x_0 = \Phi_q k_y/(2\pi H)$ is the equilibrium position of the harmonic
potential, $\omega_s = 2\vert q \vert H / (m c_0)$ is the harmonic potential
frequency, and $\epsilon_n = \vert \alpha (T) \vert - \hbar^2 k_z^2/ (2 m)
= \hbar \omega_s (n + 1/2)$ is the eigenvalue. 
The highest magnetic field at which superconductivity nucleates
occurs for $n = 0$ and $k_z = 0$ leading to the condition 
$\vert \alpha (T) \vert  = \hbar \omega_s/2$.
Isolating the magnetic field from the harmonic potential frequency 
leads to $H_{c_2} (T) = [\Phi_q/(2\pi)] 2 m \vert \alpha (T)\vert/\hbar^2$, 
which can be finally expressed in terms of the coherence length $\xi (T)$ as
$
H_{c_2} (T) 
= 
\Phi_q
/
\left[
2\pi 
\xi^2 (T)
\right].
$
Substituting $\xi (T) = \xi_{GL} \vert \epsilon (T) \vert^{-1/2}$,
we write 
$
H_{c_2} (T) 
=
H_{c_2,0} \vert \epsilon (T) \vert, 
$
where $H_{c_2,0} = \Phi_q/(2\pi \xi_{GL}^2)$.
Using again the reference field $H_{k_F} = \Phi_q k_F^2$, we obtain
the dimensionless upper critical field
$
{\widetilde H}_{c_2,0} 
= 
H_{c_2,0}/ H_{k_F}
=
1/(2\pi k_F^2 \xi_{GL}^2).
$
This expression is equivalent to the ratio 
$
\hbar \omega_{c_2,0} /(2\pi \epsilon_F),
$
where 
$
\omega_{c_2,0}
= 
q H_{c_2,0}/(mc_0)
$ is the cyclotron frequency at $H_{c_2,0}$.
In the BCS limit,
$
{\widetilde H}_{c_2,0}
=
\left[
6\pi/7\zeta(3)
\right]
\left(
T_c/\epsilon_F
\right)^2,
$
which in terms of the scattering parameter $1/(k_F a_s)$
becomes 
$
{\widetilde H}_{c_2,0}
=
D
\exp[-\pi/(k_F \vert a_s \vert )]
$
with 
$
D =
256 e^{2\gamma -4}/[7\pi \zeta(3)].
$
In the BEC limit,
$
{\widetilde H}_{c_2,0}
=
\left(
2/\pi^2
\right)
\sqrt{\epsilon_F/\vert \mu \vert}
$
which in terms of the scattering parameter $1/(k_F a_s)$
becomes 
$
{\widetilde H}_{c_2,0}
=
(2/\pi^2) k_F a_s.
$
Since $k_F \xi_{GL}$ reaches a minimum in the region near unitarity 
and $\mu = 0$, it is clear that ${\widetilde H}_{c_2,0}$ has a maximum there,
where type-II superfluids are most robust to the presence of real 
or fictitious magnetic fields. 

Before concluding, we note that
the quantum regime $\hbar \omega_c \ge 2\pi T$ (where Landau level
quantization is important) 
can be reached experimentally for $^6$Li and $^{40}$K while preserving
superfluidity.
%
%

In conclusion, we have analyzed the effects of real or fictitious 
magnetic fields during the BCS to BEC evolution of s-wave superfluids 
with direct application to ultra-cold fermionic atoms. We have shown 
that a transition from type-I to type-II charged superfluidity occurs 
as the Ginzburg-Landau paramater 
crosses its critical value $\kappa_c = 1/\sqrt{2}$ in the density 
versus interaction phase diagram of fermions of charge $q$ and 
mass $m$. We have shown that  
$^6$Li and $^{40}$K in fictitious magnetic fields 
are extreme type-II superfluids. Finally, we have indicated 
that the critical magnetic fields (real or fictitious) 
depend strongly on the scattering parameter $1/k_F a_s$ and
reach a maximum in a region near unitarity, where superfluidity 
is more robust to their penetration.

CSdM and MI would like to thank NSF (DMR-0709584) and 
ARO (W911NF-09-1-0220), and Marie Curie IRG (FP7-PEOPLE-IRG-2010-268239) 
and T\"{U}B$\dot{\mathrm{I}}$TAK, respectively, for support.


\begin{thebibliography}{99}
\bibitem{footnote} 
It may be possible that placing ultra-cold neutral fermions 
in a cavity with Raman beams could help create the screening 
necessary for the type-I to type-II transition.
\end{thebibliography}
\end{document}